\documentclass{aa}  
\usepackage{graphicx}
\usepackage{physics,amsmath}
\usepackage{txfonts}
\newcommand{\q}[1]{`#1'}
\usepackage{natbib}
\bibpunct{(}{)}{;}{a}{}{,} 
\usepackage[normalem]{ulem}

\usepackage[colorlinks=true, allcolors = blue]{hyperref}

\hypersetup{
  colorlinks   = true,
  urlcolor     = blue,
  linkcolor    = blue,
  citecolor   = blue 
}
\usepackage{xcolor}

\begin{document} 
\title{Two-dimensional, time-dependent approximate NLTE unified model atmospheres with winds for hot, massive stars}
\author{
D. Debnath\inst{1}\thanks{Corresponding author \email{dwaipayan.debnath@kuleuven.be}} 
\and J.O. Sundqvist\inst{1}
\and N. Moens\inst{2}
\and L.G. Poniatowski\inst{3}
\and C. Van der Sijpt\inst{1}
\and A.A.C. Sander\inst{4,5}
}

\institute{
Institute of Astronomy, KU Leuven, Celestijnenlaan 200D, 
3001 Leuven, Belgium
\and
Centre for Mathematical Plasma Astrophysics, Department of Mathematics, 
KU Leuven, Celestijnenlaan 200B, 3001 Leuven, Belgium
\and
Royal Observatory of Belgium, 1180 Brussels, Belgium
\and
Zentrum f{\"u}r Astronomie der Universit{\"a}t Heidelberg, Astronomisches Rechen-Institut, M{\"o}nchhofstr. 12-14, 69120 Heidelberg, \\Germany
\and
Universit\"at Heidelberg, Interdisziplin\"ares Zentrum f\"ur Wissenschaftliches Rechnen, 69120 Heidelberg, Germany}

   \date{Received XXX; Accepted XXX}

   \titlerunning{2D, time-dependent approximate NLTE unified model atmospheres with winds}
   \authorrunning{Debnath et al.}

  \abstract
   {Time-dependent, multi-dimensional unified atmosphere and wind simulations of massive stars have so far used the same mean opacities for radiative acceleration as well as heating and cooling, effectively enforcing a strong thermal coupling between gas and radiation. Although this approximation may be valid in deeper atmospheric layers, it breaks down in the extended and diluted outflowing parts. As such, in those regions the energy- and Planck-mean opacities governing the heating and cooling of the gas, $\kappa_{\mathrm E}$ and $\kappa_{\mathrm B}$, are neither the same nor equal to the flux-weighted mean opacity $\kappa_{\mathrm F}$ setting the radiative acceleration.}
   {We present an approximate NLTE procedure that accounts for scattering in the computation of $\kappa_{\mathrm E}$ and $\kappa_{\mathrm B}$ from a multitude of spectral lines in an accelerating medium, to better capture the heating and cooling in the outflowing parts of our time-dependent, radiation-hydrodynamical (RHD) simulations and present the first results for O-type stars.}
  {We evaluate $\kappa_{\mathrm F}$, $\kappa_{\mathrm E}$, and $\kappa_{\mathrm B}$ using Sobolev escape probabilities and effective thermalization parameters from an extensive line database. We calculate the RHD simulations with a hybrid opacity scheme combining Rosseland means with line opacities in an accelerating medium; however, the scattering parts are now removed for $\kappa_{\mathrm E}$ and $\kappa_{\mathrm B}$.}
  {The simulated structures originating from the iron-opacity bump below the stellar atmosphere at $\sim$$200\,$kK propagate outward and give rise to a structured line-driven outflow. High-density structures in the wind move much more slowly than their low-density counterparts. Due to their high velocity dispersion, upon interaction, they produce localized shock fronts with the gas temperature exceeding the photon temperature. The local radiative cooling time is shorter than the advection time in the densest post-shock layers, so the heated gas at the shock front remains confined to thin localized layers. For a typical O-type supergiant star with a relatively dense wind, we find that about $35\%$ of the gas is hotter than the local radiation temperature in the outflowing-wind parts of the simulation.}
  {Due to improved treatment of heating and cooling in outflowing parts, the radiation and gas temperatures in the wind of the simulated O-type star are no longer the same, as was the case in previous multidimensional unified atmosphere and wind simulations. The net result is a multicomponent wind structure not only in density and velocity, but also in temperature. This likely has important consequences for the formation and interpretation of observed O-type star wind spectra.}
      
  \keywords{ stars: massive – stars: atmospheres - stars: winds, outflows - methods: numerical - hydrodynamics - radiative transfer}

   \maketitle
   
\section{Introduction}
\label{introduction}

The interaction between radiation and matter is a dominant process governing the structure and dynamics of stellar atmospheres and winds \citep{hubeny_mihalas_book}. In a hot, massive star, radiation not only transports energy throughout the atmosphere, it also provides substantial radiative acceleration \citep{Lucy70, cak1975}. Consequently, most properties of massive stars for example, their mass-loss rates, temperatures, atmospheric structures, and observed spectra are highly influenced by the strength of such photon--gas interactions \citep{Puls24}. The deep atmosphere is collisionally dominated, and local thermodynamic equilibrium (LTE) is a reasonable description of the gas. At the photosphere and above, however, the LTE assumption breaks down due to the low densities \citep{Mihalas70}. The radiation field is increasingly diluted, collisions are inefficient, and the opacities governing the coupling between matter and radiation become influenced by the radiation field itself. In such regimes, the radiation temperature, $T_{\rm rad}$,  can deviate significantly from the gas temperature, $T_{\rm gas}$. 

Capturing opacity effects arising from the very large number of spectral lines is a classical and long-standing problem in stellar atmosphere modeling (see the detailed review in \citealt{hubeny_mihalas_book}). In a state-of-the-art stationary, 1D unified\footnote{Unified refers to a simultaneous treatment of the sub-photospheric layers and the (line-driven) wind outflow.} model atmosphere with wind codes such as \textsc{CMFGEN} \citep{hillier_1998}, \textsc{PoWR} \citep{hamann_2004}, and \textsc{FASTWIND} \citep{puls_2005, puls_2020}, the radiative transfer and the statistical equilibrium conditions are solved in full non-local thermodynamic equilibrium (NLTE). 
However, to properly compute atmospheric heating and cooling, including the "line blocking/blanketing," and to derive the acceleration including the line-driving effect, the radiative-transfer equation in such 1D codes is solved in the co-moving frame (CMF) for $\sim$ 500\,000 frequency points \citep{puls_2020} using large atomic databases consisting of several million spectral lines \citep[e.g.,][]{pauldrach_1998}. The atmospheric model is then iterated toward convergence by coupling to the NLTE equations, assuming radiative equilibrium\footnote{Alternatively, flux conservation or the equivalent assumption of thermal balance of the electrons.}, possibly also involving an iterative solution of the stationary equation of motion \citep{sander_2017, sundqvist_bjorklund_2019, krticka_2017}. However, such a brute-force approach is not computationally feasible for multidimensional (multi-D), time-dependent radiation-hydrodynamics (RHD) simulations.

In previous multi-D RHD studies of unified atmospheres and winds of Wolf--Rayet (WR) stars \citep{nico_2022b, cassie_25}, O-type stars \citep{Debnath24, asif_2025, cassie_25, lara25, nico_2025}, and luminous blue variable (LBV) stars \citep{pieter_2026}, we included line driving by means of a hybrid technique \citep{luka_2021} combining Rosseland mean opacities and line opacities derived in the \citet{sobolev_1960} approximation \citep{luka_2022, Debnath24}. The flux mean, $\kappa_{\mathrm F}$, weighs the opacity with the radiative flux and determines the radiative acceleration, and it includes both true absorption and scattering. By contrast, the energy mean, $\kappa_{\mathrm E}$, and Planck mean, $\kappa_{\mathrm B}$, control the energy transfer between gas and radiation. They determine, respectively, how efficiently the existing radiation field heats the gas and how efficiently the gas cools by thermal emission. Pure scattering transfers momentum but does not exchange energy between the radiation field and the gas; therefore, using $\kappa_{\mathrm F}$ for the energy exchange can lead to an artificially strong thermal coupling between gas and radiation. Nonetheless, in previous time-dependent, multi-D, unified simulations it has been assumed that the energy- and Planck-mean opacities are equal to the flux-mean opacity. This neglects important scattering effects by effectively forcing LTE when computing the thermal exchange between gas and radiation.
To improve this, a fast but physically motivated NLTE treatment for calculating line opacities in the outflowing parts of multi-D atmosphere and wind simulations is necessary.

In this work, we address this by developing an approximate NLTE formalism that accounts for scattering when computing the energy- and Planck-mean line opacities in the outflowing parts of the atmosphere. We calculated opacities assuming two-level atoms and using the "Munich Atomic database", which consists of more than four million spectral lines and includes 50 energy levels for elements ranging from hydrogen to zinc, with ionization up to the eighth stage \citep{pauldrach_1998, pauldrach_2001, puls_2000}. As a result, this becomes a fairly expensive computational task. To make it tractable for multi-D simulations, we introduced opacity multipliers for a grid of density, gas, and radiation temperature, and dilution factors for different Sobolev-like optical depths. We then fitted the opacity multipliers by extending the fitting formalism from \citet[building on \citealt{gayley_1995}]{luka_2022} to involve not only flux-weighted mean opacities but also the energy and Planck means. The fitted parameters were then tabulated for various densities, gas and radiation temperatures, and dilution factors. These tables can thus be interpolated during the RHD simulations to obtain the line opacity parameters based on the local conditions within the grid. We used our updated line opacities to perform the first two-dimensional (2D) O-type star simulations with this improved description of heating and cooling. In the next section, we introduce the RHD equations that were solved in our simulations. Sect. \ref{nlte_line_opacities} describes the formulation of our current formalism, with an implementation of the line opacities in our RHD simulations shown in Sect. \ref{calculation and tabulation}. 1D test models are discussed in Sect. \ref{1d_test_models}, and the first 2D simulations are discussed in Sect. \ref{2d_model}. Finally Sect. \ref{discussion} features interpretation of some of the results shown in this paper, and Sect. \ref{summary} provides a conclusion of the work and some future prospects.

\section{RHD formulation}
\label{rhd_formulation}

We considered the following RHD equations: 
\begin{align}
    & \partial_t \rho + \div{(\rho \vec{\varv)}} = 0, \label{eq:rhd1}\\
    & \partial_t (\rho \vec{\varv}) + \div{(\rho \vec{\varv} \vec{\varv} + p \tens{I}}) = -\vec{f}_\mathrm{g} + \vec{f}_\mathrm{r}, \label{eq:rhd2}\\
    & \partial_t e + \div{(e \vec{\varv} + p \vec{\varv})} = -\vec{\varv} \cdot \vec{f}_\mathrm{g} + \vec{\varv} \cdot \vec{f}_\mathrm{r} + \Dot{q}, \label{eq:rhd3}\\
    & \partial_t E + \div{( E \vec{\varv} + \vec{F}_{\mathrm{CMF}}}) = -  \tens{P} :\grad{\vec{\varv}}  - \Dot{q} .\label{eq:rhd4}
\end{align}
Here, $\vec{\varv}$ is the gas velocity vector; $\rho$ the gas density; $p\tens{I}$ the scalar gas pressure times the identity tensor, $\tens{I}$; $\tens{P}$ the radiation pressure tensor; $e$ the total gas energy density; $E$ the radiation energy density; $\vec{f}_\mathrm{g}$ the gravitational force density; $\vec{f}_\mathrm{r}$ the radiation force density; and $\vec{F}_{\text{CMF}}$ the CMF radiative flux. The equations above are closed using the ideal gas law and a flux-limited diffusion approximation following \citet{nico_2022a}. The net rate of radiative heating and cooling, $\dot{q}$, is defined as
\begin{equation}
    \dot{q} = \int_\nu \oint_\Omega \alpha_\nu (I_\nu - S_\nu ) d \Omega d \nu  
\end{equation}
where $I_\nu$ is the specific intensity at frequency $\nu$; $\alpha_\nu$ the extinction coefficient (units of inverse length), $S_\nu \equiv \eta_\nu/\alpha_\nu$ the source function for emission coefficient, $\eta_\nu$; and the integrals are over all frequencies, $\nu$, and solid angles, $\Omega$. 

For angle-independent $\alpha_\nu$ and $S_\nu$, the above may be formulated as 
\begin{equation}
      \dot{q} = 4 \pi \int_\nu \alpha_\nu (J_\nu - S_\nu ) d \nu  
      \label{eq10}
\end{equation}
where 
\begin{equation} 
    J_\nu \equiv \frac{1}{4 \pi} \oint_\Omega I_\nu d \Omega \equiv \frac{E_\nu c}{4 \pi}
\end{equation} 
is the frequency-dependent mean intensity, and, by definition, $J = \int_\nu J_\nu d \nu$, $E = \int_\nu E_\nu d \nu$, and $J = E c / (4 \pi)$, where $J$ is the frequency-integrated mean intensity and $E_\nu$ is the frequency-dependent radiation energy density. Under the assumption of source functions consisting of thermal and scattering components $S_\nu = \epsilon_\nu B_\nu + (1-\epsilon_\nu) J_\nu$, where $\epsilon_\nu$ is the relative thermal contribution, the scattering part cancels in $\dot{q}$. The Planck function $B_\nu(T_{\rm gas})$ was evaluated at the local gas temperature, and we also introduced it below $B=\int_\nu B_\nu(T_{\rm gas})\,d\nu=\sigma T_{\rm gas}^4/\pi$ as the frequency-integrated Planck function. Thus, alternatively, 
\begin{equation}
    \dot{q} = c \rho \kappa_{\mathrm E} E - 4 \pi \rho \kappa_{\mathrm B} B ,  
\end{equation}
where 
\begin{align}
    & \kappa_{\mathrm E} \equiv \int_\nu E_\nu \epsilon_\nu \kappa_\nu d \nu / E \\
    & \kappa_{\mathrm B} \equiv \int_\nu B_\nu \epsilon_\nu \kappa_\nu d \nu / B
\end{align}
are energy- and Planck-weighted mean opacities, for $\kappa_\nu = \alpha_\nu / \rho$. \\ 
For $\dot{q} = 0$, the above becomes the standard radiative-equilibrium equation often used to compute the temperature structure in stationary model atmospheres. Due to the splitting of the source function, we were able to avoid an explicit computation of the source-function-weighted opacity, $\kappa_{\mathrm S}$ \citep[e.g.,][]{hamann_grafener_2003}, which is typically required in stationary NLTE atmosphere models. In such approaches, $\kappa_{\mathrm S}$ describes the coupling between the radiation field and the local source function. In our formulation, this coupling can also be directly captured by the Planck mean opacity, $\kappa_{\mathrm B}$, for spectral lines due to our assumption of effective two-level atoms (see Sect. \ref{approx_nlte}).
    
\section{Mean line opacities in expanding atmospheres}
\label{nlte_line_opacities}

In the previous section, we introduce the flux-, energy-, and Planck-mean opacities in the radiation force and heating and cooling terms. In this section, we describe our present method of calculating and tabulating the opacities accounting for scattering in an accelerating medium using the Sobolev escape probability for all contributing lines in the database. 

\subsection{Approximate NLTE level populations}
\label{approx_nlte}

In LTE, the ionization fractions follow the Saha equation, but the diluted stellar radiation field drives departures from such an equilibrium. A correction factor can be introduced to encapsulate this departure \citep{abott_lucy_1985, Schmutz_1991, lucy_1993, Schaerer_1994, puls_2000}. Therefore, the modified ionization-equilibrium equation becomes
\begin{align}
 \frac{n_{j+1} n_{\mathrm e} }{n_j} 
        &= 2 \frac{U_{j+1}}{U_j}
 \left(\frac{2\pi m_\mathrm{e} k_\mathrm{B} T_{\rm rad}}{h^2}\right)^{3/2}
 e^{-E_{\rm ion}/k_\mathrm{B} T_{\rm rad}} \nonumber \nonumber \\
        &\quad 
 \tilde{W} \sqrt{\frac{T_{\rm gas}}{T_{\rm rad}}}\,
 \big[\zeta + \tilde{W}(1-\zeta)\big],
 \label{saha_boltzman}
\end{align}
where $n_{\mathrm e}$ is the electron density, $\tilde{W}$ is the modified dilution factor, defined in Eq.\,\eqref{Eq:dil} below (see \citealt{lucy_1993}), and $T_{\rm gas}$ and $T_{\rm rad}$ are the gas (or electron) and radiation (or photon) temperatures, respectively. $U_j$ and $U_{j+1}$ are the partition functions of the lower and upper ionization stages, and $E_{\rm ion}$ is the ionization energy. $\zeta \approx \sigma_g/\sum_l \sigma_l = 1 /\sum_l \sigma_l$ represents the fraction of recombinations that lead directly to the ground state, $\sigma_g$ is the (normalized) recombination cross-section to the ground level, and $\sigma_l$ is the cross-section for level $l$ of the same ionization stage. Following the approximate NLTE treatment adopted in \citet{abott_lucy_1985} and \citet{puls_2000}, the standard Boltzmann relation is modified to account for the dilution of the radiation field. To calculate the ionization and excitation balance, we also required the electron number density, for which we followed the procedure outlined in Appendix A.1 of \citet{luka_2022}. Our implementation of this approximate NLTE procedure has already been tested for the calculation of $\kappa_{\rm F}$ in Appendix A of \citet{sundqvist_lime}.

\subsection{Line heating and cooling}
\label{line heating and cooling}

Let us consider a spectral line with the extinction coefficient $\alpha_\nu = \alpha_l \phi_\nu$, where 
\begin{equation} 
    \alpha_l = \sigma_{\rm cl} f_{lu} n_l \left(1- \frac{n_u g_l}{n_l g_u}\right)
\end{equation} 
is the frequency-integrated line extinction and $\phi_\nu$ the line profile function, normalized such that $\int_\nu \phi_\nu\,d \nu =1$. Also, $\sigma_{\rm cl}$ is the line cross-section, and $f_{\rm lu}$ is the oscillator strength for the transition between levels $l$ and $u$. The quantities $n_l$ and $n_u$ are their respective level populations (in $\rm cm^{-3}$) with statistical weights $g_l$ and $g_u$. For this line, the net heating and cooling rate is given by (Eq.\,\ref{eq10}):
\begin{equation}
\dot{q} = \alpha_l \int_\nu \oint_\Omega (\phi_\nu I_\nu - \phi_\nu S_\nu )\, d \Omega\, d \nu, 
\label{Eq19}
\end{equation}
where we neglected interaction with the continuum. The line source function, $S_l$, varies slowly across the line profile function and is taken to be angle independent. This simplifies Eq.\,\eqref{Eq19} to
\begin{equation}
\dot{q} = 4 \pi \alpha_l \bar{J_\phi} - 4 \pi \alpha_l S_l  
\end{equation}
where $\bar{J_\phi}$ is the profile-weighted mean intensity
\begin{equation}
\bar{J_\phi} \equiv \frac{ \int_\nu \oint_\Omega 
I_\nu \phi_\nu\, d \Omega\, d \nu}{ 4 \pi}.
\end{equation}
For a two-level model atom, the line source function can be written as (e.g., \citealt[page~458]{hubeny_mihalas_book})
\begin{equation}
    S_l = \epsilon_l B_l(T_{\rm gas}) + (1-\epsilon_l) \bar{J_\phi},   
\end{equation}
where $B_l(T_{\rm gas}) = B_{\nu_0}(T_{\rm gas})$ is the Planck function evaluated at the line-center frequency, $\nu_0$, and local gas temperature, $T_{\rm gas}$, and 
\begin{equation} 
    \epsilon_l = \frac{C_{ul}}{C_{ul} + A_{ ul}} 
\end{equation}
is the line photon-thermalization probability, with $C_{ul}$ being the collisional de-excitation rate and $A_{ul}$ the Einstein coefficient for spontaneous emission. The line scattering part thus cancels out in the net heating and cooling rate, giving
\begin{equation}
\dot{q} = 4 \pi \alpha_l \epsilon_l \left(\bar{J_\phi} -B_l(T_{\rm gas})\right).   
\end{equation}
To estimate $\bar{J_\phi}$ in a medium with a significant velocity gradient, we applied the theory of \citet{sobolev_1960}, as formulated by \citet{rybicki_hummer}:
\begin{equation}
    \bar{J_\phi} = (1-\beta) S_l + \beta_{\mathrm c} I_{\mathrm c} ,
\end{equation}
where $I_{\mathrm c}$ is the illuminating specific intensity entering the line resonance zone from the continuum radiation field. Also, $\beta$ and $\beta_{\mathrm c}$ are the Sobolev escape probabilities \citep{rybicki_hummer} expressed as
\begin{align}
    & \beta = \frac{1}{4 \pi} \oint_\Omega 
    \frac{1-e^{-\tau_{\mathrm S}}}{\tau_{\mathrm S}}\, d \Omega  \mbox{\hspace{1em}} \text{and} \\
    & \beta_{\mathrm c} I_{\mathrm c} = \frac{1}{4 \pi} \oint_\Omega 
    I_c\, \frac{1-e^{-\tau_{\mathrm S}}}{\tau_{\mathrm S}}\, d \Omega\text{.}
    \label{Eq:betac}
\end{align}
Here, $\tau_{\mathrm S}$ is the Sobolev optical depth,  
\begin{equation}
    \tau_{\mathrm S} = \frac{\alpha_l c}{\nu_{\mathrm 0} |d\varv_{\mathrm s}/ds|},  
\end{equation}
where, $\varv_{\mathrm s}$ is the line-of-sight velocity. To approximate the integral in Eq.\,\eqref{Eq:betac}, we followed a customary approach for hot-star wind models and approximate $\beta_c I_c \approx \tilde{W} \beta B_l(T_{\rm rad}) $ \citep{hiller_1987, springmann_phd, puls_2000, puls_2005}. The Planck function is then evaluated at the radiation temperature $T_{\rm rad}$. The generalized dilution factor $\tilde{W}$ lies between zero and one \citep{lucy_1993}. Following \citet{lucy_1993} and \citet{springmann_phd}, we further used a spherically modified optical depth scale $\tilde{\tau}$ and the standard dilution factor (W) to set the modified dilution factor, which is given as, 
\begin{equation}
    \tilde{W} = W + \frac{3 \tilde{\tau}}{4} 
              = 0.5\left(1-\sqrt{1-\left(\frac{R_\star}{r}\right)^2}\right) + \frac{3 \tilde{\tau}}{4}\text{,}
\label{Eq:dil}
\end{equation}
This modified dilution factor $\tilde{W}$ originally stems from an analytic radiative-transfer solution for $J$  by \citet{lucy_1971} in a diluted spherical envelope illuminated by a core intensity \citep[e.g.,][]{pauldrach_1986, hiller_1987, lucy_1993, springmann_phd, puls_2000}. While approximate, it ensures a smooth connection between the geometrically diluted radiation field in the wind and the thermal radiation field in optically thick layers. In the outer wind, where optical depths are smaller, $\tilde{W}$ approaches the geometrical dilution factor $W$. In deeper atmospheric layers, the additional optical depth $\tilde{\tau}$ term in Eq.~\eqref{Eq:dil} drives $\tilde{W}$ toward unity, ensuring that the LTE limit is properly established at $\tilde{\tau} = 2/3$; for all layers of $\tilde{\tau} > 2/3$ one simply sets $\tilde{W} = 1$ by definition. With these approximations, we obtain the profile-weighted mean intensity: 
\begin{equation}
    \bar{J_\phi} = (1-\chi_l) \tilde{W} B_l(T_{\rm rad}) + \chi_l B_l(T_{\rm gas}),   
\end{equation}
where 
\begin{equation}
    \chi_l = \frac{\epsilon_l (1-\beta)}{\epsilon_l(1-\beta) + \beta}
\end{equation}
acts as an effective thermalization parameter for interactions within the Sobolev resonance zone. 
Even for small values of $\epsilon_l$, we can have $\chi_l \rightarrow 1$ if $\beta \ll \epsilon_l$, implying that a strong, optically thick line may undergo several interactions before it escapes, hence increasing the probability of thermalization. Finally, we can obtain the heating and cooling rates as 
\begin{equation}
    \dot{q} =  4 \pi \alpha_l \epsilon_l (1- \chi_l) \left( \tilde{W} B_l(T_{\rm rad}) -B_l(T_{\rm gas}) \right). 
    \label{Eq:qdot_line}
\end{equation}
This indicates that the line heating and cooling depend on the temperature imbalance between the diluted stellar radiation field and the local gas temperature, which is modulated by the efficiency of photon thermalization within the resonance zone. The factor $\epsilon_l$ in Eq.~\eqref{Eq:qdot_line} provides the probability that an interaction with the line results in a thermal energy exchange with the gas, rather than merely scattering the photon. The additional factor $\chi_l$ accounts for the fact that photons trapped within an optically thick Sobolev resonance zone are absorbed and re-emitted many times before they ultimately escape; if each of these interactions has $\epsilon_l > 0$ the net effect is to increase the likelihood for local thermalization beyond the local value of $\epsilon_l$. Overall, Eq.~\eqref{Eq:qdot_line} shows that a line contributes to the heating and cooling when it has a nonzero thermal component and the local radiation field differs from the gas thermal component. The line heats the gas if the diluted radiation field provides more energy to the system than the gas can emit away at the line frequency.

\subsection{Formulation of energy-, Planck-, and flux-mean line opacities}
\label{fomulationn_of_energy}

Following the notation introduced in the previous section, Eq.\,\eqref{Eq:qdot_line} can be rewritten as
\begin{equation}
    \dot{q} = c \rho \kappa_{\mathrm E} E - 4\pi \rho \kappa_{ \mathrm B} B,  
\end{equation}
where $\kappa_{\mathrm E}$ and $\kappa_{\mathrm B}$ describe the total energy- and Planck-mean opacities, and their corresponding opacity multipliers are described in Eq. \eqref{eq37} (note that $\kappa_{\mathrm E} = \kappa_{\mathrm J}$ in this formalism). For an individual line, $l$, the corresponding contributions are
\begin{align}
    \kappa_{E, l} &= \kappa_{\mathrm 0} q_l \epsilon_l (1-\chi_l) w_{E,l}, 
    \label{kappa_E} \\
    \kappa_{B, l} &= \kappa_{\mathrm 0} q_l \epsilon_l (1-\chi_l) w_{B,l}.
    \label{kappa_B}
\end{align}
Here, $w_{B,l}$ and $w_{E,l}$ are line illumination functions for the Planck- and energy-mean opacities, respectively. We adopted the notation of \citet{luka_2022}, which presented an analogous illumination function in the context of flux-weighted line opacities (see Eq. 18 in \citealt{luka_2022} for the exact definition). However, contrary to the flux mean, the Planck mean was weighted using $T_{\rm gas}$, and the energy mean was weighted using the diluted $T_{\rm rad}$. The line strength, $q_l$, is defined as 
\begin{equation}
    q_l = \frac{\alpha_l}{\nu_{\mathrm 0} \kappa_{\mathrm 0} \rho},
\end{equation}
where $\kappa_{\mathrm 0}$ is a scaling opacity, $\nu_{\mathrm 0}$ is the line-center frequency, and $h$ is the Planck constant. $B \equiv \sigma T_{\rm gas}^4/\pi$ is the frequency-integrated Planck function, and $E = 4\sigma T_{\rm rad}^4 / c$ is the frequency-integrated radiation energy density. Summing this over an ensemble of (intrinsically non-overlapping) spectral lines with the index $i$ finally yields, for effective energy and Planck line opacity multipliers,
\begin{equation}
   M_{\mathrm [E,B]} \equiv \frac{\kappa_{\mathrm  [E,B]}}{\kappa_{\mathrm  0}} = \sum_i q_i \epsilon_i (1- \chi_i) w_{[E,B],i}.  
   \label{eq37}
\end{equation}
We here extended the opacity-multiplier formalism previously used for the flux mean in \citet{luka_2021} (LTE) and in \citet{sundqvist_lime} (approximate NLTE, Appendix A). For the calculations of these opacity multipliers, we used the same characteristic optical-depth scaling and fitting routine (see below) as in \citet{luka_2021}, but this time deriving separate multipliers for the energy and Planck means. In comparison to the flux-weighted opacity, the expression in Eq.\,\eqref{eq37} involves a combination of $\epsilon$ and $\chi$ in the contributions to the energy and Planck mean\footnote{Note that in the flux mean defined by \citet{luka_2022} the Sobolev optical depth is captured directly by the Sobolev escape probability entering that expression. In Eq.\,\eqref{eq37} it is instead accounted for in the $\chi$ parameter.}. A pure scattering line with $\epsilon_i = \chi_i = 0$ thus does not contribute to $\dot{q}$. Contrary to that, a thermalized optically thin line ($\beta_i \rightarrow 1$) with $\epsilon_i = 1$ and $\chi_i \rightarrow 0$ contributes a factor of $q_i w_{[E,B],i}$, analogous to the flux-weighted line opacity. In our new formalism, this flux-weighted mean opacity can be computed exactly as in \citet{luka_2021}, however, this time explicitly using $T_{\rm rad}$ in the corresponding weighting functions $w_{F,i}$ (see Appendix A in \citealt{sundqvist_lime}). 
\begin{figure}
    \centering
    \includegraphics[width=1\linewidth]{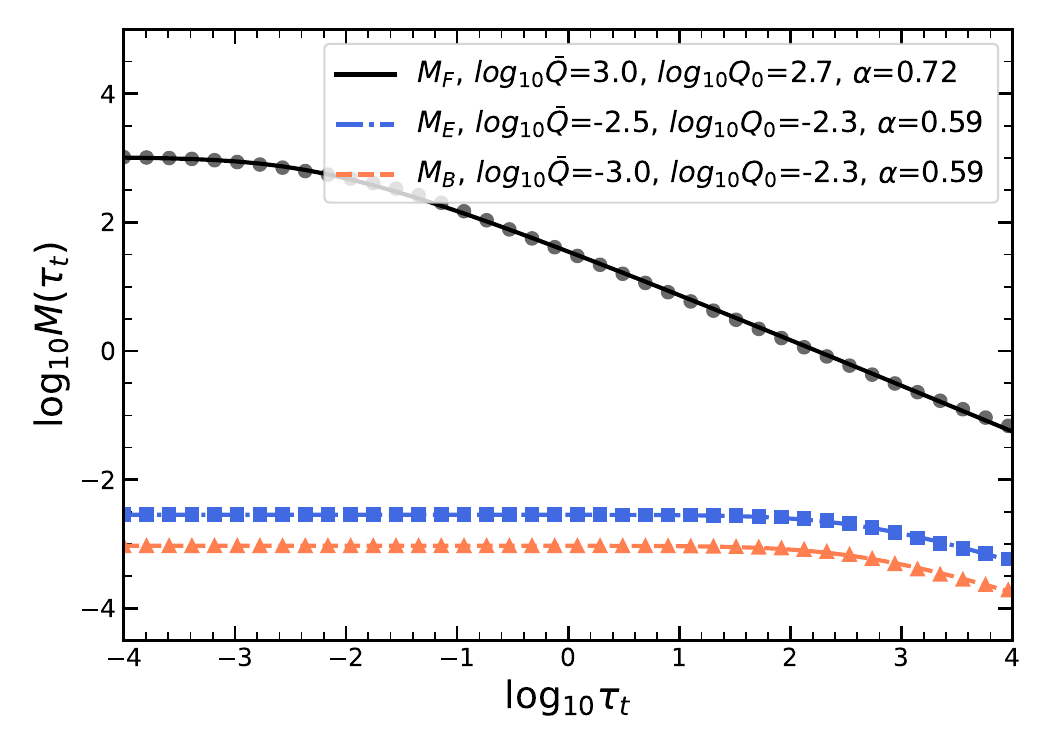}
    \caption{Opacity multipliers as a function of the characteristic scale, $\tau_{t}$, for the flux, energy, and Planck mean at a radiation temperature of $T_{\rm rad}$ = 40 kK, density $log_{10} \rho \, {\rm [g/cm^3]}$ = -13, dilution factor $W = 0.33$, and the temperature ratio $T_{\rm gas}/T_{\rm rad}$ = 0.8. The solid lines provide the fits to the opacity multiplier using the fitting formalism introduced by \citet{gayley_1995}.}
    \label{fig:mt}
\end{figure}

\section{Calculation and tabulation of line mean opacities} 
\label{calculation and tabulation}

To compute the Sobolev-based line-mean opacities, we used the Munich Atomic Database \citep{pauldrach_1998,pauldrach_2001,puls_2000}, which contains $\sim$4 million spectral lines from hydrogen up to zinc. The data include ionization stages up to the eighth stage (ions that are seven times ionized). For the energy levels, the database provides degeneracies and excitation energies for every present ionization stage. For every spectral line, the database provides the $gf$-value, the wavelength, and the upper and lower levels for that transition. Building on the notation suggested by \citet{gayley_1995} and subsequently used by \citet{luka_2022}, we determined opacity multipliers ($M_{\mathrm [F,E,B]}$) by fitting the corresponding line opacity parameters $\alpha$, $Q_{\mathrm 0}$, and $\bar{Q}$ with an analytical fit function described in \citet{luka_2022} and \citet{sundqvist_lime}. $\bar{Q}$ gives the saturation value of the opacity multiplier in the optically thin limit, $Q_{\mathrm 0}$ provides an effective maximum line strength, and $\alpha$ is a power-law index describing a mixture of optically thick and thin lines. As seen in Fig.~\ref{fig:mt}, the opacity multiplier associated with the flux-mean opacity, $M_{\mathrm F}$, has significantly higher numerical values than the energy-mean opacity multiplier, $M_{\mathrm E}$, and Planck-mean opacity multiplier, $M_{\mathrm B}$. This arises because, unlike $M_{\mathrm F}$, the energy and Planck opacity multipliers $M_{\mathrm E}$ and $M_{\mathrm B}$ (and hence their associated opacity means) receive no contribution from scattering. As the photosphere of an O-type star is typically located at the characteristic scale  ${\tau_{\mathrm t}} =\tau_{\mathrm s}/q_l \simeq 10^{2 \dots 3}$ \citep{luka_2022,sundqvist_lime}, the flux-mean opacity multiplier remains significantly below its $\bar{Q}$ value, whereas the Planck and energy means are already in their optically thin limit.

For use in the hydrodynamical calculations, these parameters were calculated for a four-dimensional grid in density, radiation temperature, gas-to-radiation temperature ratio, and dilution factor. The Sobolev-based line opacities can be reconstructed using the tabulated fit parameters and the parameter $\tau_{\rm t}$ at any point in the grid. Examples of this method for the flux-mean opacity can be found in \citet{luka_2022}, \citet{nico_2022b} and \citet{Debnath24}. We limited our radiation temperature to $\log_{10} T_{\rm rad} \in [4,5]$ because of the ionization stages included in our atomic database. As the photosphere of an O-type star is around $T_{\rm rad} \sim 30-40$ kK, deeper layers with $T_{\rm rad} > 10^5$ K will be very optically thick, and thus Sobolev-based opacities vanish there anyway. Since our tables are discrete, we performed a quadrilinear interpolation scheme when calculating the opacity multipliers within the actual RHD simulations. 

\begin{figure*}
 \centering
 \includegraphics[width=18cm]{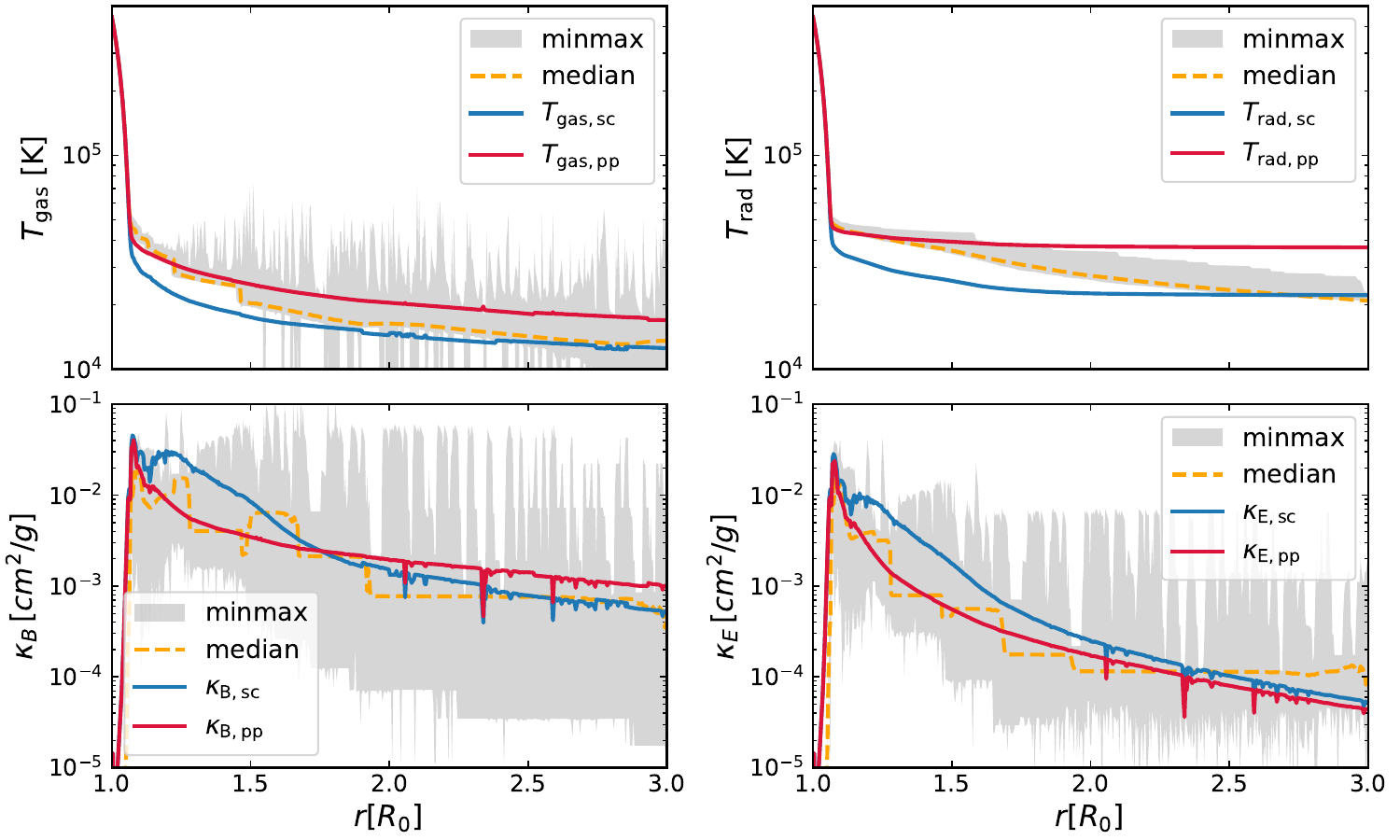}
      \caption{Comparison of temporal median, computed from relaxed snapshots (dashed orange line) of the 1D RHD simulation (Sect. \ref{1d_test_models}), with the corresponding quantities obtained under the assumption of stationary radiative equilibrium. The red lines show the quantities calculated using the plane-parallel approximation (Eq.~\ref{Eq:diff_pp}), while the blue lines represent the results including spherical correction terms. The corresponding quantities are annotated with the subscripts "pp" or "plane-parallel" and "sc" or "spherical correction", respectively. From left to right, the panels display the gas and radiation temperatures, followed by the Planck-mean and energy-mean opacities. The gray shaded regions indicate the distribution of data points from the snapshots.}
  \label{quantities}
\end{figure*}
    
\section{1D test models} 
\label{1d_test_models}

In this section, we explore the line opacity method described above in 1D RHD simulations based on the setup in \citet{nico_2022a} but with parameters suitable for O-type stars. 

\subsection{Time-dependent simulation}
\label{time_dependent_sims}

We solved the time-dependent RHD equations on a finite volume mesh as described in \citet{nico_2022b} and \citet{Debnath24}. However, we focus on the radial spatial dimension for our first test models in this section. The partial differential equations (PDEs) outlined in Sect. \ref{rhd_formulation} were solved using the RHD module \citep[][of MPI-AMRVAC \citealt{xia_2018, rony_2023}]{nico_2022a}. 
We used a hybrid opacity technique \citep{luka_2021} that incorporates both the Rosseland means \citep{opal} and the above-described effects from line opacity in accelerating media. However, contrary to our previous simulations \citep{nico_2022b,Debnath24,asif_2025,cassie_25,lara25,nico_2025, pieter_2026} we do not assume the flux-, energy-, and Planck-mean opacities to be the same. The line opacities are instead computed as outlined above, removing scattering parts from the heating and cooling rates. The Rosseland mean opacities controlling the deeper atmospheric structure consist of various interactions; however, scattering does not influence the thermal coupling between gas and radiation. Accordingly, we approximately removed the influence of scattering by subtracting the electron (Thomson) continuum scattering contribution. Accordingly, the potential effects of line scattering and blanketing are still neglected in these regions. The iron-opacity peak at around $150\,\mathrm{kK}$ can significantly enhance the Rosseland mean opacity in deep subsurface layers. In 2D and 3D simulations, this produces large-scale density and velocity structures that can generate variability up to the photosphere and the wind \citep{Debnath24, cassie_25, lara25, nico_2025}. For such O-type stars, \citet{cassie_25} find that the growth of these structures is not compatible with unstable gravity modes but could be through the Rayleigh-Taylor instability. In effectively 1D radial simulations, or generally in simulations where motions cannot develop laterally, the same opacity enhancement instead tends to produce strongly inflated and highly fluctuating atmospheres \citep{asif_2025}. To avoid such a behavior in the present 1D test case, whose purpose is to illustrate the thermal decoupling between gas and radiation, we restricted the opacity sources to just electron scattering and line driving, essentially meaning $\kappa_{\mathrm E} \approx \kappa_{\mathrm B} \approx 0$ in the deep atmospheric layers where the Sobolev optical depth is high (pure scattering atmosphere). We stress that this simplification is not applied to the 2D simulations presented in later sections. Recent work by \citet{Key_2025} shows that a steep atmospheric stratification can self-excite oscillations (dominating around the Lamb cutoff frequency) through the Sobolev line-force dependence on the velocity gradient in layers just beneath the photosphere. Such Lamb oscillations are more prevalent in cases with steep density gradients at the photosphere-to-wind transition, particularly in models with low classical Eddington ratios ($\Gamma_\mathrm{e}$). For the present 1D test case, we therefore adopted a comparatively higher $\Gamma_{\mathrm e}$ than usual for a typical O-type supergiant star (OSG), to reduce this variability and obtain a quasi-steady reference model for comparing $T_{\rm gas}$, $T_{\rm rad}$, $\kappa_{\mathrm E}$, and $\kappa_{\mathrm B}$.  However, such oscillations may still arise in multi-D simulations, but they are typically smoothed out due to multi-D averaging effects. 

Figure~\ref{quantities} illustrates a representative 1D test simulation for a star with $M_\star/M_\odot = 58.28$, $R_\star/R_\odot = 14.47$, $\Gamma_\mathrm{e} = 0.67$, and $T_{\rm eff} = 50\,\mathrm{kK}$. The dashed orange curve displays the median\footnote{The median provides a more robust estimate than the mean in these 1D simulations, as it is less sensitive to numerical fluctuations.} for the radiation and gas temperatures and $\kappa_{\mathrm E}$ and $\kappa_{\mathrm B}$ for a set of relaxed snapshots. The gray shading represents the distribution (region between minimum and maximum value) of temperatures or the opacities within those snapshots. However, as we move further away from the stellar photosphere, the radiation field becomes increasingly diluted. The energy- and Planck-mean opacities, $\kappa_{\mathrm E}$ and $\kappa_{\mathrm B}$, differ substantially from $\kappa_{\mathrm F}$ and from each other; this is because they explicitly depend on the thermalization properties of the lines for both thermal means and the local radiation field in the case of $\kappa_{\mathrm E}$. As a result, the gas and radiation temperatures already decouple near the photosphere. However, we note that the LTE $\kappa_{\mathrm F}$ agrees reasonably well with the $\kappa_{\mathrm F}$ calculated with the current methodology, in the region where the mass loss is initiated close to the photosphere; the resulting change in the mass-loss rate is negligible (see Appendix \ref{flux_mean_appendix} and the discussion in Appendix A of \citealt{sundqvist_lime}).

\subsection{Comparison to stationary radiative equilibrium temperature structure} 
\label{rad_eqv}

To further benchmark the time-dependent simulations, we constructed a simplified radiative-equilibrium iterative model, which 

(i) starts with initial guesses for $B$ and $J$; 

(ii) computes mean line opacities according to our current method outlined above;  

(iii) uses the radiative equilibrium $\dot{q} =0$ i.e., $\kappa_{\mathrm E} J = \kappa_{\mathrm B} B$ condition to obtain a new estimate of $B$ (and thereby $T_{\mathrm gas}$) and 

(iv) computes a flux-limiter $\lambda$ analogous to the FLD description used in the time-dependent simulations, and updates $J$ (i.e., $T_\mathrm{rad}$) from the stationary radiative-transfer moments assuming plane-parallel geometry (Eq. \ref{Eq:diff_pp}) and with corrections for spherical geometry 
(see Appendix A in \citealt{nico_2022a}) for direct comparison with the time-dependent RHD simulations. For the plane-parallel case, we thus solve
\begin{align}
     & -\frac{d}{d\tau} \left(\lambda \frac{dJ}{d\tau}\right) + \tilde{\epsilon}_{\mathrm J} J 
     = \tilde{\epsilon}_{\mathrm B} B, 
    \label{Eq:diff_pp} 
\end{align}
where $\tilde{\epsilon}_{\mathrm J} = M_{\mathrm E}/(1+M_{\mathrm F})$, $\tilde{\epsilon}_{\mathrm B} = M_{\mathrm B}/(1+M_{\mathrm F})$, and $d\tau = -\kappa_{\mathrm 0}(1+M_{\mathrm F})\rho dr$. This procedure was iterated until the mean opacities, as well as $J$ and $B$, converged. Figure~\ref{quantities} shows the resulting radiation and gas temperatures, computed assuming plane-parallel ($T_{\rm gas, pp}$ and $T_{\rm rad, pp}$) and with the spherical correction ($T_{\rm gas, sc}$ and $T_{\rm rad, sc}$) approximations. The same acronyms also represent the opacities calculated for the plane-parallel ($\kappa_{E, \mathrm{pp}}$ and $\kappa_{B, \mathrm{pp}}$) and spherically corrected ($\kappa_{E, \mathrm{sc}}$ and $\kappa_{B, \mathrm{sc}}$) equilibrium models.

The temperatures computed under the plane-parallel approximation ($T_{\rm gas, pp}$ and $T_{\rm rad, pp}$) agree well with both the median radiation and gas temperatures of the 1D simulation near the lower boundary. However, the plane-parallel approximation fails to match the corresponding values from the 1D RHD simulation at larger radii. In contrast, when spherical correction terms are included in Eq.\,\eqref{Eq:diff_pp}, the resulting gas ($T_{\rm gas, sc}$) and radiation ($T_{\rm rad, sc}$) temperatures show good agreement with the median values of the 1D simulation at larger distances, as illustrated in Fig.~\ref{quantities}.
The remaining discrepancies here may arise from our treatment of the correction terms in the RHD simulations themselves and will require some further investigation in future work.

Additionally, in our stationary radiative-equilibrium model, all opacities were computed directly from the complete discrete sum (Eq.\,\ref{eq37}), without relying on fitting functions or interpolation from discrete opacity tables. As is evident from Fig.~\ref{quantities}, the opacities obtained via interpolation from discrete line-opacity parameters agree reasonably well with the self-consistently calculated values, lending support to our fit function and interpolations. It should be noted that there are deviations by factors of a few, but we emphasize that these opacity differences do not stem from inaccuracies in the fitting or interpolation procedure itself, they mainly arise due to the residual differences in $T_{\rm gas}$ and $T_{\rm rad}$ between the stationary model and the 1D RHD simulation as discussed above. Nevertheless, we can infer that the interpolation and fitting scheme captures the main opacity trends well, while the discrepancy is inherited from the imperfect agreement in the aforementioned thermal structure.

\section{2D O-type star model}
\label{2d_model}

\begin{figure*}
 \centering
 \includegraphics[width=18cm]{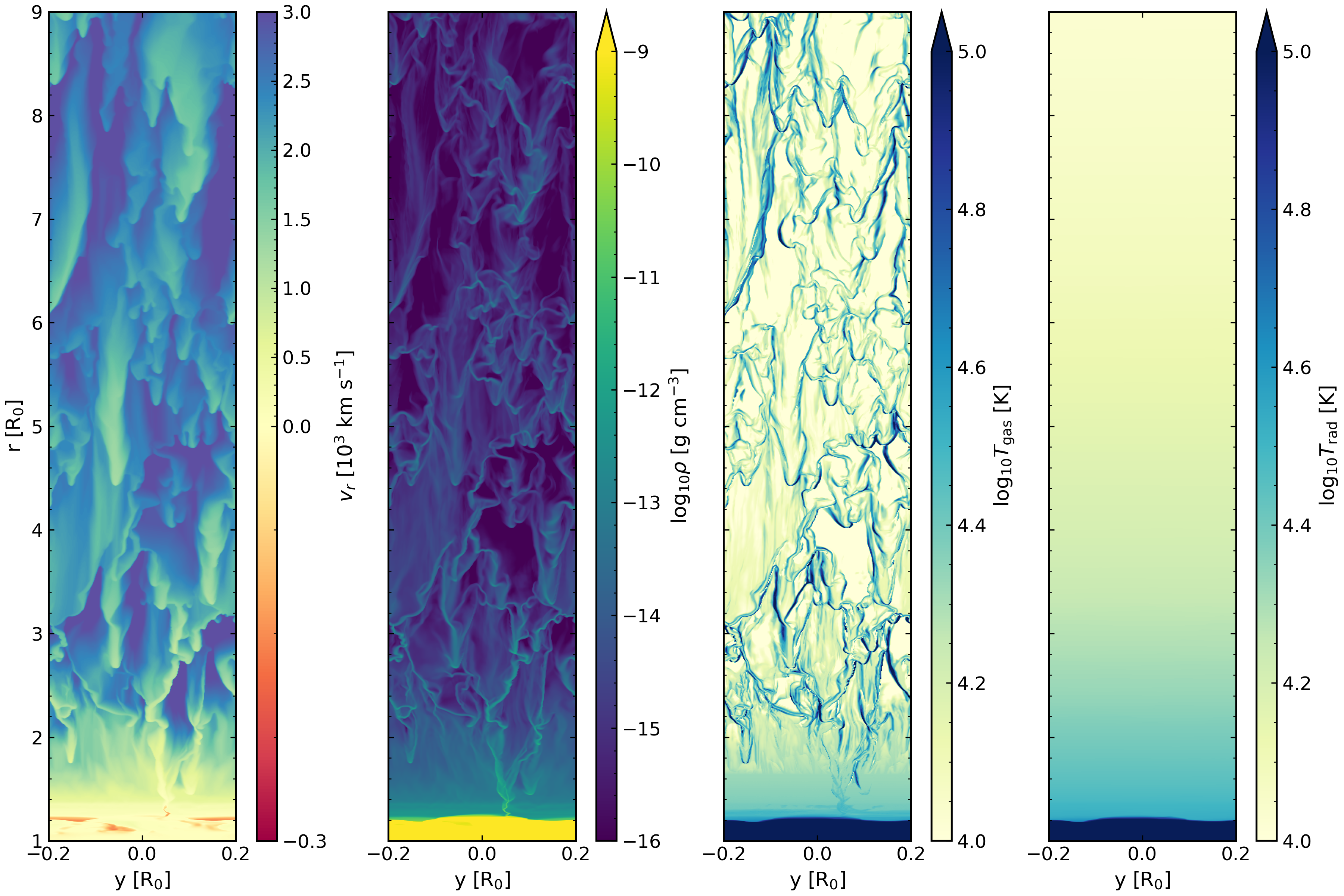}
      \caption{Color maps of the radial velocity, density, gas, and radiation temperature (from left to right, respectively) after the simulation has relaxed from its initial conditions. }
  \label{summary_grid}
\end{figure*}

We next applied our method to a 2D simulation and explored the effects of the presently calculated Sobolev-based line opacities on temperature structures in the wind. The 2D models were calculated in a time-dependent manner, solving the RHD equations outlined in Sect.~\ref{rhd_formulation} on a Cartesian grid, but incorporating spherical corrections in the divergence terms \citep{sundqvist_2018, nico_2022a}. The grid spans from $r = R_0$ to $r = 9R_0$ in the radial direction, and $ 0.4R_0$ in the lateral direction. The lower boundary was placed at a temperature of 500 kK, allowing us to capture the stable radiative zone, the iron-opacity bump \citep{iglesias_1992, opal} at around 150 kK, the photosphere, and the outflowing region. This setup ensured that we properly resolved both the iron-opacity bump and the associated turbulence that develops from it \citep{Debnath24, cassie_25}. The grid resolves down to $2 \times 10^{-3}R_0$ in both radial and lateral directions throughout the whole box (covering five or more pressure scale heights). The initial and boundary conditions were constructed following \citet{Debnath24} and \citet{nico_2025}. To account for the geometric effects of the stellar disk when calculating the flux-, energy-, and Planck-mean opacities in the wind, we incorporated a finite disk-correction factor as described in \citet[and references therein]{Debnath24}.

Figure~\ref{summary_grid} shows a color map of the radial velocity, density, gas, and radiation temperature of a snapshot after the model has relaxed for a star with a mass of $M_\star/M_\odot = 58.28$, classical Eddington of $\Gamma_\mathrm{e} = 0.32 $, radius of $R_\star/R_\odot = 17.39$, luminosity of $\log_{10} \left(L/L_\odot\right) = 5.85  $, effective gravity of $\log \ g = 3.73 $, effective temperature of $T_{\rm eff} [\mathrm{kK}] = 40.3 $ and mass-loss rate of $\log_{10} \dot{M} [ M_\odot\,\mathrm{yr}^{-1}]= - 5.55$.

\subsection{Structure formation, shock heating and de-coupling of gas and radiation temperature}
\label{structure_formation}

Previous studies on the atmospheres of O-type stars have shown that structure originates deep within the iron-opacity region and propagates outward toward the photosphere and the outflowing line-driven wind \citep{Jiang_2015, Debnath24, cassie_25,nico_2025,lara25}, resulting in a turbulent atmosphere. However, those studies focused mainly on the iron-opacity bump and the transition from the photosphere to the onset of the supersonic outflow. As such, the outflowing wind was not fully resolved, unlike the deeper regions and the photosphere. Contrary to those studies, in this work we also resolved the wind to the same extent as the deeper regions. From Fig.~\ref{summary_grid}, it can be seen that the atmospheric structures generated in the deeper layers also give rise to a structured outflow. It can be seen that in the outflowing line-driven wind region, relatively high-density clumps are moving much more slowly than the background rarefied medium, with relative differences of $\sim 1000 \, \rm km/s$ or more. Due to the inverse density dependence of line force, the rarefied medium experiences a much greater force due to line driving than the higher density clump, giving rise to this velocity dispersion (see \citealt{nico_2022b} and \citealt{Debnath24} for a more detailed explanation). This results in the formation of a shock front at the boundary between the two media. While similar shock configurations were already present in earlier multi-D simulations, their thermal impact could not be captured, since strong thermal coupling between the gas and the radiation field was enforced by assuming $\kappa_{\mathrm E} = \kappa_{\mathrm B} = \kappa_{\mathrm F}$. In the present work, the separation of the flux-, energy-, and Planck-mean opacities enable a better treatment of radiative heating and cooling in the wind, such that shock-heated gas is no longer forced to cool instantaneously.  The gas temperature can now decouple from the radiation temperature. In regions of strong compression, such as shock fronts, this leads to sustained heating of the gas. As such, for quite a few of these clumps, at the shock fronts the gas is heated to temperatures approaching $10^5\,\mathrm{K}$. This results in two distinct gas and radiation-temperature structures in our simulations. 

\begin{figure}
 \centering
 \includegraphics[width=1\linewidth]{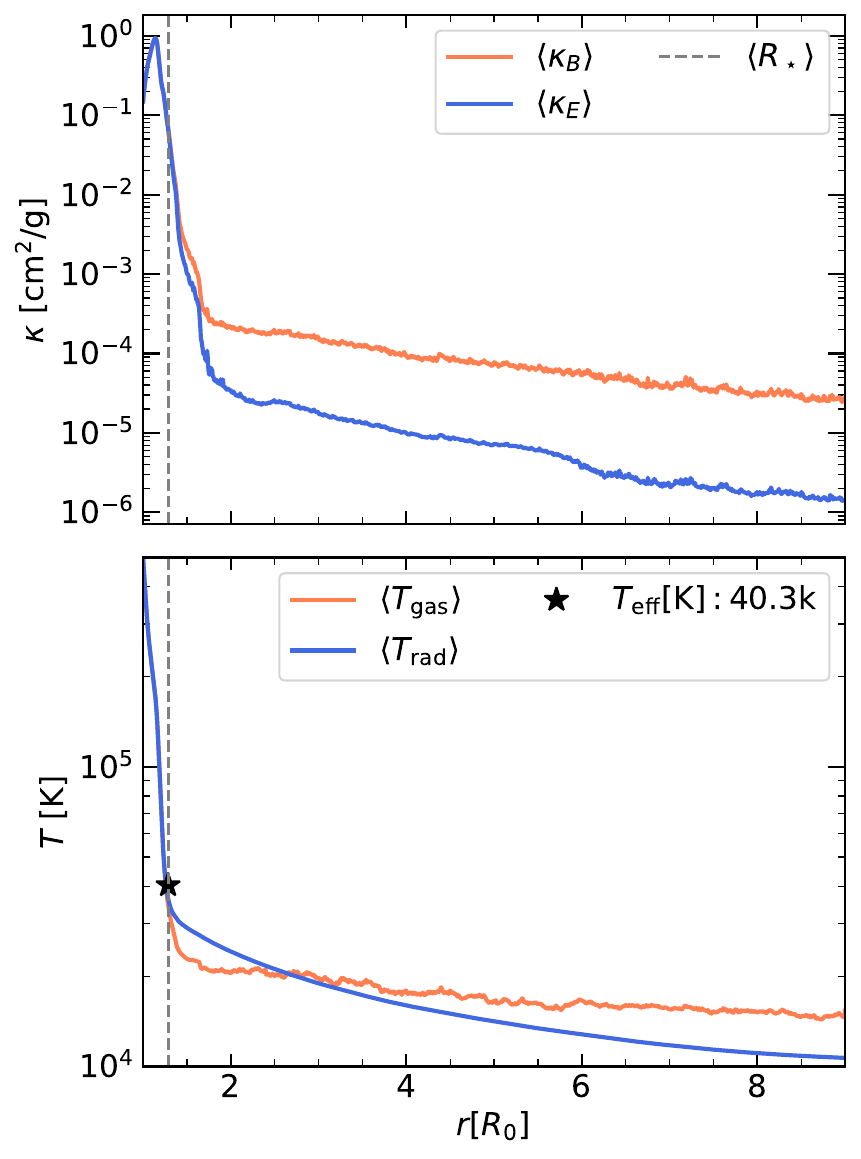}
      \caption{Laterally and temporally averaged (for relaxed snapshots) Planck and energy mean (top) and gas and radiation temperatures (bottom) for the 2D simulation.}
  \label{temp_kappa}
\end{figure}

Figure~\ref{temp_kappa} shows the laterally and temporally averaged gas and radiation temperatures, together with the energy- and Planck-mean opacities. Similar to the 1D test case shown in Fig.~\ref{quantities}, the numerical value of the Planck-mean opacity is larger than the energy-mean opacity as the radiation becomes increasingly diluted further away from the star. Since the Planck mean governs radiative cooling, this should result in the gas being cooler than the radiation, as observed for the 1D simulation in Sect. \ref{1d_test_models}. However, for the 2D simulation, the gas remains, on average, hotter than the radiation temperature in the wind (see Fig.~\ref{temp_kappa}). As discussed above, at the shock fronts, the gas can be heated to temperatures significantly exceeding the local radiation temperature. When averaged, these localized but intense heating events compensate for the enhanced radiative cooling, causing the average gas temperature in the wind to remain higher than the radiation temperature. To further quantify the amount of hot gas in our simulation, we computed the volume occupied by the hot gas, $V^{\rm T}_{\rm ff}$, in the outer wind regions, which we define as the domain beyond $3 R_0$. We find that approximately $35\%$ of the gas, on average, is heated to temperatures beyond the local radiation temperature, whereas about $4\%$ is heated to temperatures above the effective temperature of the star. This substantial fraction of shock-heated gas further explains why the average gas temperature remains higher than the radiation temperature further out in the wind. \citet{lagae_2021} found in their line-deshadowing instability \citep[LDI;][]{owocki_1984} simulations of an OSG model that about \(7\%\) of the gas remains hot. This is similar to the fraction of gas in our model that reaches temperatures above the effective temperature. We note, however, that the characteristic temperatures of LDI-induced shocks are significantly higher than those found here.

\subsection{Optically thin cooling}
\label{optically_thin}

\begin{figure}
    \centering
    \includegraphics[width=1\linewidth]{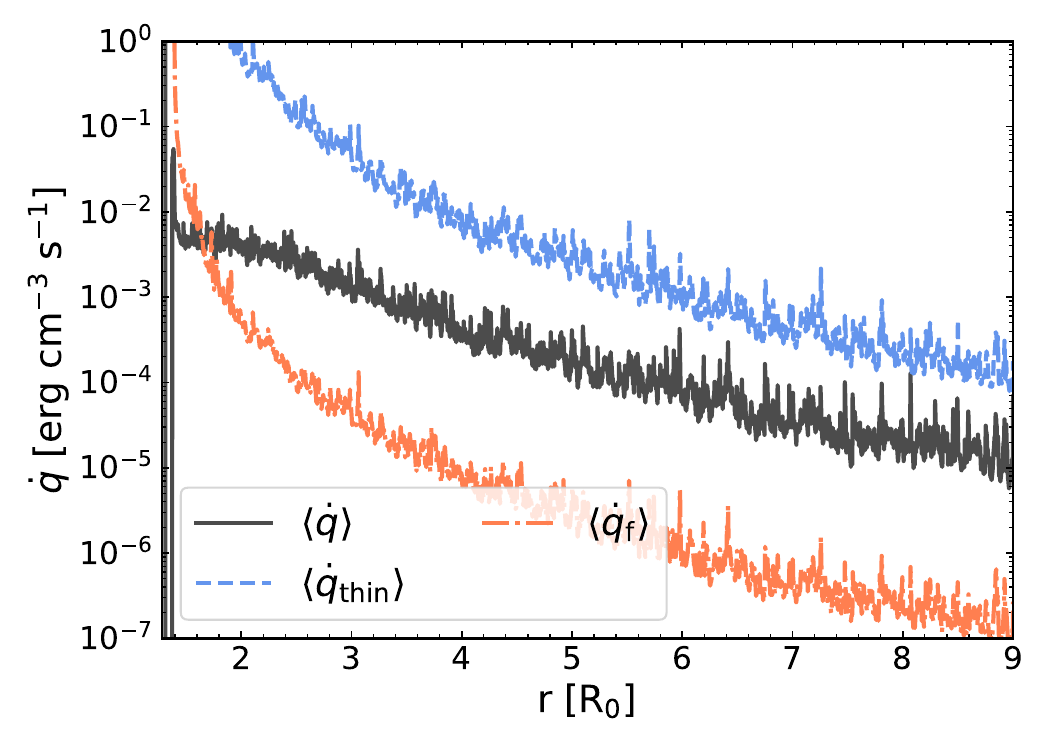}
    \caption{Comparison of laterally and temporally averaged net heating and cooling rate $\dot{q}$ (in black) with the optically thin cooling rate (in blue) calculated for the same set of snapshots, only showing the parts beyond the photosphere. The orange curve shows the analytical fit adopted from \citet{feldmeier_1997}; see text.}
    \label{qdot_comparison}
\end{figure}

We compare the net radiative-heating-and-cooling rate from our current Sobolev treatment with the optically thin cooling prescriptions commonly used in hydrodynamical simulations. It is important to note that these two approaches are based on different physical assumptions, as further discussed below. Furthermore, the atomic databases underlying our comparisons are not the same, making direct one-to-one comparisons challenging. Nonetheless, our comparisons highlight that radiation-field-dependent treatment of the thermal balance can deviate from prescribed optically thin cooling curves for certain cases and should be taken into consideration. As previously discussed, we find that in many layers of our simulations, the Sobolev-based line-opacity means $\kappa_{\mathrm E}$ and $\kappa_{\mathrm B}$ have already reached their optically thin limits. This arises because in the calculation of the Sobolev-based part of $\kappa_{\mathrm E}$ and $\kappa_{\mathrm B}$ only interactions that involve pure absorption, and therefore exchange energy between the gas and radiation, are taken into account. The gas becomes transparent to thermalizing photons much more rapidly than for $\kappa_{\mathrm F}$, which also includes contributions from scattering. As a result, the transition from the optically thick, exponential regime of the line opacity multiplier associated with $\kappa_{\mathrm F}$ occurs at a significantly lower characteristic depth than for $\kappa_{\mathrm E}$ and $\kappa_{\mathrm B}$ (see Fig. ~\ref{fig:mt}). This allowed us to compare our heating and cooling formalism to the optically thin cooling tables used in \textsc{AMRVAC} \citep{schure_2009}. These tables were calculated assuming collisional ionization equilibrium with the \textsc{SPEX} package for a range of temperatures for solar abundances from \citet{anders_1989}. It provides the integrated emissivity $\Lambda(T)$ normalized per $n_{\mathrm e} n_{\mathrm H}$. The dependence of the individual ion densities $n_i$ is already present in the tabulated cooling function, and the volumetric cooling rate is
\begin{equation}
    \dot{q}_{\rm thin}(T) = -\,n_{\rm e} n_{\rm H}\,\Lambda(T).
\end{equation}
The electron density $n_{\mathrm e}$ depends on the actual composition of the ions; \citet{schure_2009} also provided $n_{\mathrm e}/n_{\mathrm H}$ as a function of temperature, 
\begin{equation}
    \dot{q}_{\rm thin}(T) = -\,n_{\rm H}^2\, \frac{n_{\rm e}}{n_{\rm H}}\Lambda(T)= -\,n_{\rm H}^2\,\Lambda_{\rm hd}(T) .
\end{equation}
This allowed us to read the electron density rather than explicitly solving for the ionization balance. Fig.~\ref{qdot_comparison} compares this optically thin cooling to the $\dot{q}$ obtained using our methodology, averaged laterally and temporally, focusing on regions $r > R_\star$.  As seen in Fig.~\ref{qdot_comparison}, there is a systematic offset of about an order of magnitude between our predicted radiative cooling rate and the optically thin cooling rates. This discrepancy may arise from several factors. First, our $\dot{q}$ is calculated using the Sobolev approximation and assuming two-level atoms using a large atomic line database, accounting for both line heating and cooling, whereas the optically thin formalism only accounts for cooling.  In that formalism, heating from the parent star is often mimicked by introducing a floor temperature (e.g., \citealt{lagae_2021}). Notably, our approach does not rely on such assumptions. A further difference is that these cooling tables are typically calculated under the assumption that the plasma has reached collisional ionization equilibrium (CIE), which is only expected to hold when the plasma is collision-dominated. Additionally, photoionization is neglected when calculating such rates. In the atmospheres and outflows of massive stars, however, the stellar radiation field can significantly alter the ionization structure and should be taken into account. \citet{schure_2009} also note that the deviations from CIE become significant for shocks with velocities above 150 $km \, s^{-1}$. Indeed, we find that the largest differences between $\dot{q}$ and $\dot{q}_{\rm thin}$ are at the shock fronts. Therefore, these cooling tables may not, in general, be well-suited for calculating radiative cooling timescales for moderately heated gas in the atmospheres and outflows of massive stars. On the other hand, our present treatment only includes bound--bound processes with ionization stages that are up to seven times ionized, meaning that if the gas is heated to above $T_{\rm gas} \sim 10^5$ K, our database does not have the appropriate ionization stages. Presently, if $T_{\rm gas} > 10^5$ K in our simulations, we linearly extrapolated from the tabulated values at $T = 10^5$ K.

Figure~\ref{qdot_comparison} also shows the modified optically thin cooling rate, $\dot{q}_{\rm f}$, calculated using the fitting formula proposed by \citet{feldmeier_1997}, with a cutoff temperature of $T_{\rm swi}$ = 15 $\langle T_{\rm eff} \rangle$, a cooling function normalization of $A = 1.6 \times 10^{-22} \mathrm erg \, cm^3 \, s^{-1}$, and a power-law exponent \(\delta = 2\) below the cutoff. This fitting formula was introduced to better reproduce the observed X-ray properties of O-type stars, by circumventing the numerical "overcooling" problem that often occurs in optically thin cooling treatments \citep{feldmeier_1997} due to advective diffusion at shock fronts \citep{lagae_2021} and the oscillating cooling instability \citep{langer1981,langer1982}. In the following section, we discuss how the temperatures required to reproduce the observed X-ray emission were not reached in our simulations. At present, it remains unclear whether this primarily reflects numerical diffusion in our models, similar to the effects discussed by \citet{lagae_2021}, or whether the shocks produced in our simulations are simply too weak as compared to those generated in LDI simulations.

\section{Discussion} 
\label{discussion}

\begin{figure*}
    \centering
    \includegraphics[width=1\linewidth]{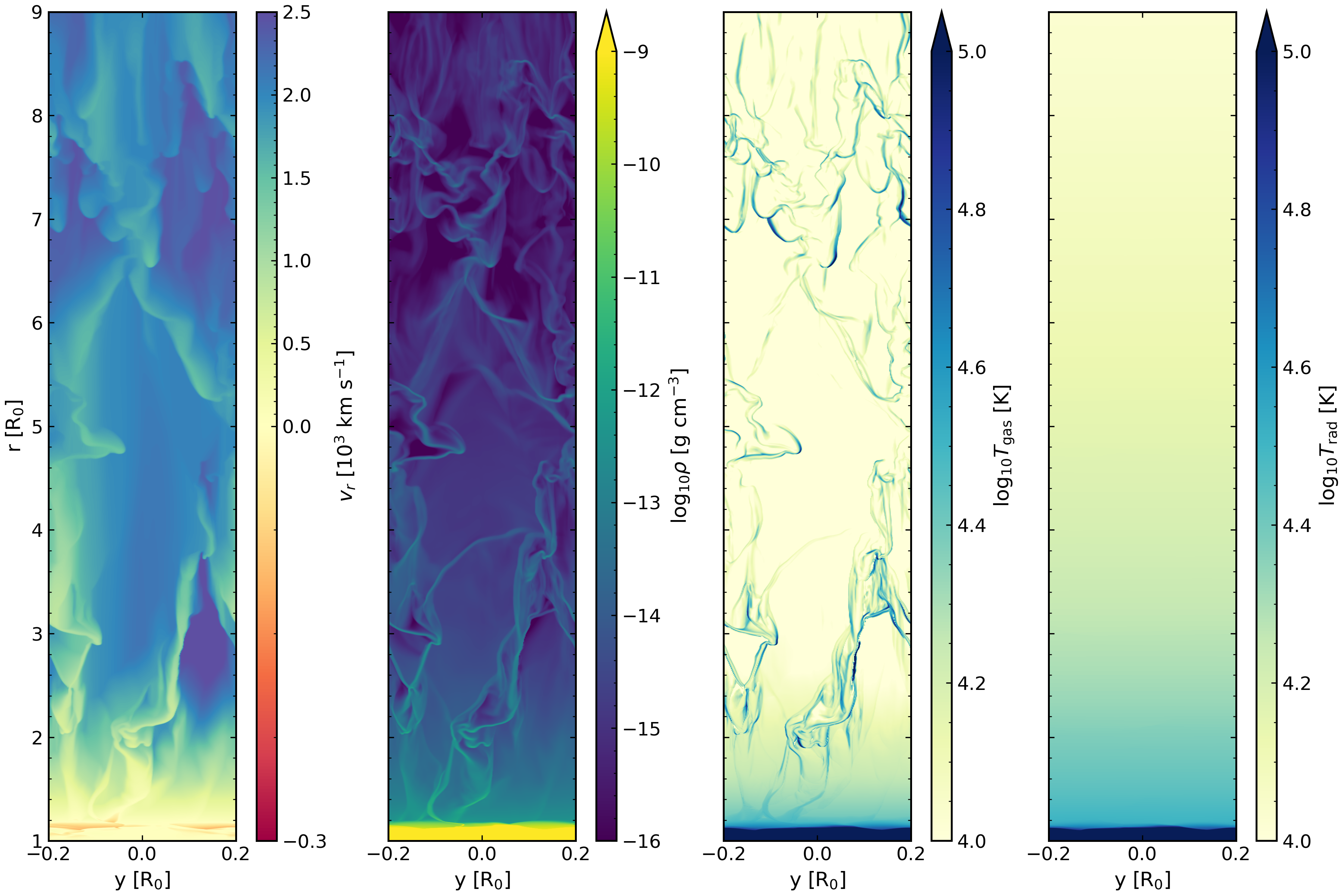}
    \caption{Color maps of radial velocity, density, gas, and radiation temperature (from left to right, respectively) for a stereotypical O-type dwarf star.}
    \label{odwarf_fig}
\end{figure*}

\subsection{Emergence of a multi-temperature structure}
\label{two_temperature}

A key result of this work is the emergence of a multi-temperature structure in multi-D atmosphere and wind simulations of hot, massive stars. In deep atmospheric layers, as the density increases, collisions become more frequent, leading to thermalization of many spectral lines ($\epsilon_l \rightarrow 1$). As we move toward the photosphere and beyond, the density decreases, and collisions become less efficient. The radiation field becomes nonlocal and is increasingly diluted further away from the star, and the net radiative heating of the gas decreases. This produces the difference in gas and radiation temperature visible in Figs. \ref{quantities} and \ref{temp_kappa} and marks the onset of the two-component gas and radiation-temperature structure in our simulations. 

In the wind, the interaction between the fast, rarefied gas and the slow, dense clumps gives rise to strong, localized shocks. At the shock front, the gas is compressed and heated. The compressional heating rate per unit volume of gas is $\dot{Q}_{\rm comp} = - p_{\rm g} \nabla\cdot\vec{\varv}$. At the shock front, $\dot{Q}_{\rm comp}$ becomes positive, heating the gas compressively. At the same location, the radiative sink term is also significant and, in fact, dominates the energy balance. This is also reflected in the corresponding timescales; the net radiative timescale, as shown in Eq. \eqref{trad}, is at least an order of magnitude smaller than the compressional heating timescale in the post-shock region. It should be noted that we are concerned primarily with the shock-front regions, where radiative losses are strong, and the net radiative term is dominated by cooling, essentially making $\dot{q} \simeq \dot{q}_{\rm cool} = \rho c a_{\mathrm r} \kappa_{\mathrm B} T_{\rm gas}^4 $. The kinetic energy deposited at the shock is thus efficiently radiatively cooled before it can be advected out. Consequently, the shock-heated regions remain spatially confined, appearing as thin layers in the flow.

We define the net radiative timescale as 
\begin{equation}
    t_{\rm rad} = \varepsilon/\left |\dot{q} \right| = \frac{\varepsilon}{\left |\rho c a_{\mathrm r} (\kappa_{\mathrm E} T_{\rm rad}^4 - \kappa_{\mathrm B} T_{\rm gas}^4) \right |}
    \label{trad}
\end{equation}
where $\varepsilon = e - \rho \varv^2/2 $ is the internal gas energy. We define the advective timescale as, 
\begin{equation}
    t_{\rm adv} \equiv R_\star/\varv_{\rm eff}
    \label{tadv}
\end{equation}
with $R_\star$ being the laterally and temporally averaged photospheric radius, and $\varv_{\rm eff}$ the radial velocity at that grid point. $t_{\rm rad}$ measures how quickly a shocked gas parcel can radiatively adjust the thermal energy it gained at the shock, whereas $t_{\rm adv}$ measures the time required for that parcel to be advected over a characteristic length here taken as one stellar radius. Also, at the shock front, where radiative cooling dominates over radiative heating, $t_{\rm rad} \ll t_{\rm adv}$ and the shock-heated gas cools down before it can travel downstream. As a result, the shock-heated gas stays confined to narrow spatial regions behind the shock front rather than forming an extended hot component in the outflow. This is because the compressed gas at the shock front is much denser than the surrounding rarefied medium. As the gas is heated locally at the shock front, its internal energy increases roughly as $\varepsilon \propto \rho T_{\rm gas}$. However, once the gas temperature significantly exceeds the radiation temperature, $T_{\rm gas} \gg T_{\rm rad}$, the net radiative loss rate is given by approximately $\dot q \propto \rho\,\kappa_{\mathrm B} T_{\rm gas}^4$. The radiative cooling rate then increases sharply with this increase in gas temperature, resulting in a shorter radiative cooling timescale. 

Our work demonstrates that once the enforced equality $\kappa_{\mathrm E}=\kappa_{\mathrm B}=\kappa_{\mathrm F}$ is relaxed, embedded wind shocks give rise to hot, localized regions where gas temperatures are significantly higher than radiation temperatures. In this sense, the present simulations provide a physically motivated origin for hot gas in line-driven winds. However, it should be noted that the temperatures required for the formation of X-rays ($\sim 1 \rm MK$ and above) are not produced in our current simulations, which is likely due to the absence of LDI, as our method does not capture the line deshadowing instability. Alternatively, this may partly result from numerical diffusion affecting the smallest-scale structures in our simulations as discussed previously.

\subsection{Comparison to a low-$\Gamma_\mathrm{e}$ O-type dwarf star model}
\label{Odwarf}

The formation of structure in our simulations is positively correlated with the proximity of the star to the Eddington limit. Stars closer to the Eddington limit develop stronger subsurface turbulence \citep{Debnath24, cassie_25, nico_2025} and more pronounced density contrasts, which are then amplified by line driving in the wind and vice versa. 

We calculated a 2D O-type dwarf star simulation that is further away from the Eddington limit with a mass of $16 M_\odot$, Eddington factor of $\Gamma_\mathrm{e} = 0.21$, surface gravity of $\log g \simeq 4 $, and effective temperature of 42\,kK. Consequently, the sub-photospheric structure generation is much less pronounced, which gives rise to a lower number of shock fronts in the outflowing regions. As evident from Fig.~\ref{odwarf_fig}, the majority of the gas remains in close equilibrium with the radiation temperature, with only $14\%$ of the gas heated to temperatures above the local radiation temperature in the low-density outflowing regions, and an even smaller fraction of the gas approximately $ 0.8 \%$ heated to temperatures higher than the stellar effective temperature. 

In contrast, \citet{lagae_2021} performed LDI simulations, including optically thin radiative cooling of the shock-heated gas, of both O-type dwarf and supergiant stars; the authors found that in the dwarf case nearly $70\%$ of the gas in the wind is heated to temperatures above $10^5\,\mathrm{K}$. The high percentage of hot gas in the O-type dwarf star model was attributed to strong shocks caused by the LDI and low wind density, resulting in inefficient cooling (at least when applying the correction to the cooling curves suggested by \citealt{feldmeier_1997}; see discussion above). On the other hand, these LDI simulations assume line driving from a smooth radiative surface, thus neglecting any turbulence generated in subsurface regions. This highlights two distinct origins of wind structures. In the case of OSGs close to the Eddington limit, the outflow consists of a combination of subsurface turbulence and wind instabilities. By contrast, for O-type dwarfs further away from the Eddington limit, structure formation in the wind will likely be dominated by LDI. 

Additionally, \citet{Puls_2009} suggested that deep-seated clump formation can lead to subsequent collisions between wind clumps, generating strong X-rays. A possible scenario may thus be that subsurface structures provide perturbations that can seed and enhance the LDI effect. Overall, strong localized shocks therefore appear to be an unavoidable feature of hot-star winds. Although the resulting shock temperatures may vary, such shocks can be triggered by subsurface turbulence, the LDI, or an interaction between the two. Overall, we can conclude that strong localized shocks are unavoidable in hot-star winds. While the associated shock temperature might differ, both subsurface turbulence and the LDI (or the combination of both) can trigger them.

\subsection{Limitations and future improvements}
\label{limitations}

\paragraph{Frequency-dependent radiation temperature.} In addition to bound--bound transitions, the thermal coupling between the atmospheric plasma and photons is also influenced by bound-free processes, such as the hydrogen and helium continuum edges. Since photoionization cross-sections typically scale as roughly $\sigma_{\rm bf}\propto \nu^{-3}$  \citep{Biberman_1962, puls_2005} above the threshold energy level for that transition, these edges introduce strong frequency dependence in the continuum opacity. Consequently, different frequency bands decouple from the gas at different optical depths. A natural extension would be to include bound--free and free--free continuum opacities and to generalize the radiation field to a number of frequency bins, which would be especially important to compute high-energy emissivities stemming from the shock-heated gas as well as the line-blanketed pseudo-continuum radiation field necessary for realistic NLTE post-processing of multi-D RHD simulations \citep{lara25}.

\paragraph{Fitting the line-opacity multiplier globally.} The slope of the line opacity multiplier can vary as the dominant contributing element or ionization stage to the line opacity changes with Sobolev-like optical depth. 
When applied globally, this fitting formula can lead to a mismatch between the fitted opacity multiplier and the "true" value at a given Sobolev-like optical depth. To mitigate such issues, \citet{sundqvist_lime} restricted the fit to optical depths up to a critical value prevalent at the wind's critical point relevant for setting the (stationary) mass-loss rate for a particular star. This strategy is not viable for our multi-D simulations, however, where the optical depth depends on position and time in the simulation.  Work is currently underway in our group to replace the fitting routine with a machine-learning approach that predicts the line-opacity multiplier directly, thereby avoiding fit functions and similar associated inconsistencies.

\paragraph{LDI.}
One major limitation in our current method is that the line-driving opacities are computed using a Sobolev approach, which assumes that the hydrodynamical conditions are constant over a resonance zone. As discussed above, line driving in hot-star winds is highly unstable due to the LDI effect. LDI has its strongest response near the Sobolev length ($l_{\rm sob} = \varv_{\rm th}/d\varv_s/ds$) \citep{owocki_1984}, but perturbations on substantially larger scales can also grow and steepen into strong shocks over a few stellar radii \citep{feldmier_2020}. In our simulations, extensive structures in the wind arose even with Sobolev-based line driving because of the turbulent wind-launching region stemming from instabilities associated with the iron bump. As discussed in Sect. \ref{Odwarf}, LDI likely plays an important role in producing shock-heated gas in OB dwarfs, and it might also further affect wind clumping properties of denser winds. However, it remains both methodologically and computationally very challenging to include the LDI in a multi-D model also covering the subsurface iron-opacity bump region (see \citealt{puls_owocki_1996} for the substantial challenges involved in even 1D time-dependent LDI simulations from a smooth stellar surface).

\section{Summary and conclusion}
\label{summary}

In this work, we developed and implemented a formalism to treat approximate NLTE effects and scattering in a multi-D, time-dependent, RHD, unified atmosphere and wind model of a hot, massive O-type star. Our new method separates the flux-, energy-, and Planck-mean opacities, allowing a better description of radiative acceleration and of the net heating and cooling of the gas in the optically thin outflowing regions of the atmosphere. More specifically, we derived a mean opacity description accounting for scattering in a medium with a significant velocity gradient. We then precomputed line opacity multipliers for a grid of density, radiation temperature, gas-to-radiation temperature ratio, and dilution factor, and fitted them using the distribution function concepts of \citet{gayley_1995} and \citet{luka_2022}. These line opacity parameters were then used to compute the line contributions to $\kappa_{\mathrm E}$, $\kappa_{\mathrm B}$, and $\kappa_{\mathrm F}$ while running the multi-D O-type star models using \textsc{MPI-AMRVAC}. 
Enforcing the energy- and Planck-mean opacities to match the flux-weighted mean opacity, as done in our previous multi-D RHD simulations, artificially couples gas and radiation temperatures in the wind, thereby underestimating the shock heating of the gas. Our new treatment of the energy balance here leads to local pockets in the wind where the gas temperature significantly exceeds the local radiation temperature. The presence of shock-heated gas strongly suggests that thermal decoupling naturally contributes to the occurrence of high-ionization wind lines and the general complexity observed in the wind spectra of O-type stars. The simulations presented here thus represent first steps toward self-consistent multidimensional O-type star wind models with realistic multicomponent density, velocity, and temperature structures. Such models will be key for future spectroscopic analyses aimed at reproducing the complex spectra of O-type stars without invoking ad hoc fitting parameters.

\begin{acknowledgements}
    The computational resources used for this work were provided by Vlaams Supercomputer Centrum (VSC), funded by the Research Foundation-Flanders (FWO) and the Flemish Government. DD, JS, NM, CVdS acknowledge support from the European Research Council (ERC) under the Horizon Europe grant agreement 101044048, from the Belgian Research Foundation Flanders (FWO) Odysseus program under grant number G0H9218N, from FWO grant G077822N, and from KU Leuven C1 grant BRAVE C16/23/009. AACS acknowledges support by the Deutsche Forschungsgemeinschaft (DFG, German Research Foundation) in the form of an Emmy Noether Research Group – Project-ID 445674056 (SA4064/1-1, PI Sander) and a DFG research grant - Project-ID 496854903 (SA4064/2-1, PI Sander). This project was co-funded by the European Union (Project 101183150 - OCEANS). We thank the referee Prof. Achim Feldmeier for constructive comments on the manuscript. The authors also acknowledge the fruitful discussions, suggestions, and comments from the present and past members of the EQUATION team. Data analysis and visualization in this project relied on NumPy \citep{harris_2020}, SciPy \citep{virtanen_2020}, and Matplotlib \citep{hunter_2007}.

\end{acknowledgements}

\bibliographystyle{aa}
\bibliography{references_ostar} 

\begin{appendix}

\section{Comparison of the flux-mean opacity between the LTE and 
NLTE tables}
\label{flux_mean_appendix}

As discussed previously, the flux-mean opacity in the region where mass loss is initiated is very similar to that obtained under the LTE assumption. This has also been noted by \citet{sundqvist_lime}, who show that near the critical point, where mass loss is initiated, the dilution factor is not yet sufficiently low to cause a significant departure from LTE. 

\begin{figure*}
 \centering
 \includegraphics[width=18cm]{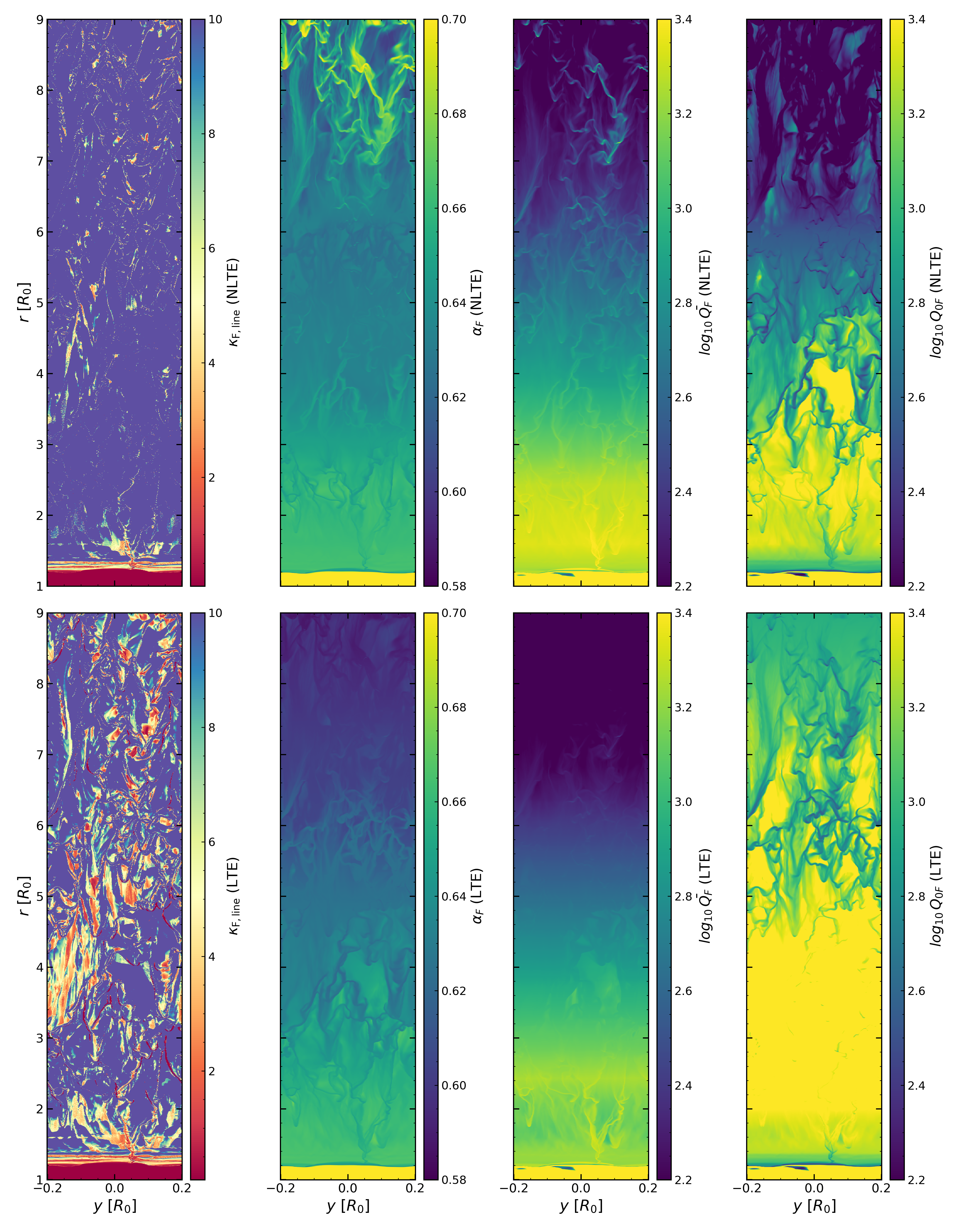}
      \caption{Color maps of the flux-mean opacity, the line force parameter $\alpha_{F}$, $\bar{Q}_F$, and  $Q_{0F}$ for the same snapshot shown in Fig.~\ref{summary_grid}. The top panel shows the values computed for the methodology discussed in this work, and the bottom panel displays the corresponding quantities assuming LTE.}
  \label{kappaf}
\end{figure*}

However, further out in the wind, the radiation field becomes significantly diluted. As a result, the ionization structure increasingly shifts from that predicted under LTE, leading to changes in the line-force parameters, particularly $\alpha_F$, which is highly sensitive to such shifts in the ionization balance. This is clearly seen in the second panel of Fig.~\ref{kappaf}, where the approximate-NLTE values of $\alpha_F$ show significant deviations in the outer regions compared to those computed under the LTE assumption. In contrast, the parameter $\bar{Q_F}$ appears largely unaffected by these changes in ionization structure. However, $Q_0$ shows a substantial difference between the NLTE and LTE cases. This behavior is expected; as $\alpha_F$ increases, the optical depths at which the force multiplier reaches its maximum value shift to larger optical depths, resulting in lower $Q_{0F}$ values in the outer wind for the NLTE case.

For the O-type stars considered in this work, these effects do not appear to significantly alter the overall wind dynamics. This is also illustrated in Fig.~\ref{kappaf}, the top-left panel shows the flux-mean opacity (only the Sobolev-enhanced line part), here, $\kappa_{F, line}$ computed using the current approach, while the bottom-left panel shows the corresponding LTE results. As evident from the figure, the values of $\kappa_{F, line}$ agree closely between the LTE and the present throughout the domain, with only minor deviations in a few localized regions.

However, in other regimes, such as BSGs, the impact may be substantial. For instance, the \q{cool-plate} models from \citet{pieter_2026} assume fixed line-force parameters throughout the domain. This simplification was adopted to ensure that the launched wind escapes the simulation domain. However, a change in the dominant ionization stage could increase the line force enough to allow the wind to escape in cases where it would otherwise remain only marginally unbound. A detailed investigation of these effects is beyond the scope of the present work and will be addressed in future studies.

\end{appendix}

\end{document}